\documentclass[useAMS,usenatbib]{mn2e_rev}
\usepackage{graphicx}
\usepackage{stackrel}
\newcommand{\hem}{\hspace{1em}}
\newcommand{\hen}{\hspace{0.5em}}
\newcommand{\rmd}{{\rm d }}
\newcommand{\rme}{{\rm e }}
\newcommand{\msun}{{\mathsf M}_{\odot}}
\newlength{\oldparindent}
\title[Linear Potential]{Schwarzschild and linear potentials in \\
Mannheim's model of conformal gravity}

\author[Peter R. Phillips]{Peter R. Phillips* \\
Department of Physics, Washington University, St. Louis, MO 63130 \\
\\
\rm Accepted 2018 May 14. Received 2018 May 13; in original form 2018 March 5
}

\begin{document}

\maketitle

\begin{abstract}
We study the equations of conformal gravity, as given by Mannheim, in the
weak field limit, so that a linear approximation is adequate. Specialising
to static fields with spherical symmetry, we obtain a second-order equation
for one of the metric functions. We obtain the Green function for this
equation, and represent the metric function in the form of integrals over
the source. Near a compact source such as the Sun the solution no longer
has a form that is compatible with observations. We conclude that a
solution of Mannheim type (a Schwarzschild term plus a linear potential
of galactic scale) cannot exist for these field equations. 
\end{abstract}




\section{INTRODUCTION}
\label{sec:intro}

In this paper we will derive solutions, in the weak field limit, of the
field equations of conformal gravity as given by \citet[equation 186]{mann6};
(this paper will be referred to as PM from now on):
\begin{equation}
4 \alpha_g W^{\mu \nu} \equiv 4 \alpha_g \left[ W^{\mu \nu}_{(2)}
- \frac{1}{3} W^{\mu \nu}_{(1)} \right] = T^{\mu \nu}\,.
\label{eq:field1}
\end{equation}
The tensor $W^{\mu \nu}$ is derived by variation of the Weyl action, defined
in PM (182). Its two separate parts,
$W^{\mu \nu}_{(1)}$ and $W^{\mu \nu}_{(2)}$, are defined in
PM (107) and (108); these definitions are repeated here:
\begin{eqnarray}
W_{(1)}^{\mu \nu} & = &
2 g^{\mu \nu} \left( R^{\alpha}_{\;\;\alpha} \right)^{;\beta}_{\;\;;\beta}
- 2 \left( R^{\alpha}_{\;\;\alpha} \right)^{;\mu;\nu}
\nonumber \\
  & & {}
- 2 R^{\alpha}_{\;\;\alpha} R^{\mu \nu}
+ \frac{1}{2} g^{\mu \nu} \left( R^{\alpha}_{\;\;\alpha} \right)^2
\nonumber \\
W_{(2)}^{\mu \nu} & = & \frac{1}{2} g^{\mu \nu}
\left( R^{\alpha}_{\;\;\alpha} \right)^{;\beta}_{\;\;;\beta}
+ R^{\mu \nu ; \beta}_{\hem \hen ;\beta}
- R^{\mu \beta ; \nu}_{\hem \hen ;\beta}
\nonumber \\
  & & {} - R^{\nu \beta ; \mu}_{\hem \hen ;\beta}
- 2 R^{\mu \beta} R^{\nu}_{\;\;\beta}
+ \frac{1}{2} g^{\mu \nu} R_{\alpha \beta} R^{\alpha \beta} \,.
\end{eqnarray}

$\alpha_g $ in (\ref{eq:field1}) is a dimensionless coupling constant. (We
adopt the notation of \citet{wein2}, with units such that
$c = \hbar = 1$.) 

The energy-momentum tensor, $T^{\mu \nu}$, is derived from an action
principle involving a scalar field, $S$ (see PM (61)). Appropriate variation
of this action yields $T^{\mu \nu}$ as given in PM (64). In Mannheim's
model, the solutions of the field
\linebreak

\setlength{\parindent}{0pt}
{\small *\ E-mail: prp@wuphys.wustl.edu \\
\hspace{2in}\\
\copyright\  2018 The Author(s)\\
Published by Oxford University Press on behalf\\
of the Royal Astronomical Society}

equations undergo a symmetry breaking
transition (SBT) in the early Universe, with $S$ becoming a constant, $S_0$.
Making this change in PM (64) we obtain
\begin{eqnarray}
T^{\mu \nu} & = & -\frac{1}{6} S^{2}_{0} \left( R^{\mu \nu}
- \frac{1}{2} g^{\mu \nu} R^{\alpha}_{\;\;\alpha} \right)
\nonumber \\
  & & {} - g^{\mu \nu} \lambda S_{0}^{4} + T_{M}^{\mu \nu}\,,
\label{eq:mat1}
\end{eqnarray}
where $T^{\mu \nu}_{M}$ is the matter tensor, containing all the usual
fermion and boson fields. From now on we will ignore the term
$g^{\mu \nu} \lambda S_{0}^{4}$, because we are not concerned with the
Hubble expansion.
\setlength{\parindent}{\oldparindent}

We break from Mannheim's development at this point. The factor $1/6$ in
(\ref{eq:mat1}) derives from the original, conformally invariant action. A
SBT, however, will not in general preserve such relations, and we will
instead write
\begin{eqnarray}
T^{\mu \nu} & = & -4 \alpha_g \eta \left( R^{\mu \nu}
- \frac{1}{2} g^{\mu \nu} R^{\alpha}_{\;\;\alpha} \right)
+ 4 \alpha_g \xi T_{M}^{\mu \nu}\,,
\label{eq:mat2}
\end{eqnarray}
so that the field equations can be written
\begin{equation}
W^{\mu \nu} + \eta \left( R^{\mu \nu}
- \frac{1}{2} g^{\mu \nu} R^{\alpha}_{\;\;\alpha} \right) 
= \xi T_{M}^{\mu \nu}\,.
\label{eq:field3}
\end{equation}
$\xi$ is dimensionless, but $\eta$ has dimension ${\rm length}^{-2}$, so
its magnitude can be written $|\eta | = 1/r_{0}^{2}$, where $r_0$ divides
lengths into two regimes, in one of which ($r < r_0 $) $W^{\mu \nu}$ is
dominant, and in the other ($r > r_0 $) the Einstein tensor,
$R^{\mu \nu} - g^{\mu \nu}R^{\alpha}_{\;\;\alpha} /2$.

We will call these equations the Weyl-Einstein equations, or
``W-E equations'' for short. We will not try to justify these equations;
Mannheim has written extensively in support of them. We are concerned only
with some of their consequences.

In the important special case that $\alpha_g W^{\mu \nu}$ is
negligible, or even identically zero, we obtain equations of Einstein form:
\begin{equation}
R^{\mu \nu} - \frac{1}{2} g^{\mu \nu} R^{\alpha}_{\;\;\alpha}
= \frac{\xi}{\eta} T_{M}^{\mu \nu}\,.
\end{equation}
If $\xi / \eta = -8 \upi {\rm G}_0$, where ${\rm G}_0$ is the usual Newtonian
gravitational constant, we regain the usual Einstein
equations, as given, for example, in \citet[equation 16.2.1]{wein2}.

In the opposite limit, $\eta \rightarrow 0$, we obtain
\begin{equation}
W^{\mu \nu} = \xi T_{M}^{\mu \nu}\,,
\label{eq:bach}
\end{equation}
the Bach equations. Some solutions of these have been obtained by
\citet{fiesch}.

We can take the trace of (\ref{eq:field3}), to get
\begin{equation}
R^{\alpha}_{\;\;\alpha} = -\frac{\xi}{\eta} T^{\alpha}_{\;\;\alpha}\,,
\label{eq:trace}
\end{equation}
which is, of course, the same as we would get from the Einstein equations
since $W^{\mu \nu}$ is traceless.

From this Mannheim derives a traceless energy-momentum tensor,
PM (65). We shall not use this, however, because it contains less
information than the original tensor, and must be supplemented by the
trace equation.

No exact solutions of the W-E equations seem to be available,
except for the usual Schwarzschild solution, which satisfies both the
Einstein and the Bach equations independently.

In this paper we will seek a solution of the W-E equations in the limit of
weak fields. This should be adequate for studies of
galactic rotation and gravitational lensing, and may give us insight into
what a more complete solution would look like. We will be particularly
interested in the Solar System (SS), for which the most incisive
observations exist. We are trying to construct a theory that is similar
to Mannheim's, so we will choose values of parameters that seem likely to
bring this about.

We can locate the present paper within the context provided by several
papers about conformal gravity, both critical and supportive, that have
appeared in recent years. \citet{mann7} has responded to the critique of
\citet{flanagan}; this debate will not concern us here. Several papers,
\citep{edpar,walker,yoon}, have shown that a linear potential of Mannheim
type is incompatible with observations; in the present paper we show that a
linear potential is not a consequence of the the field equations of
conformal gravity, so these critiques are not needed.

\citet{gegen2017} have used fourth-order gravity to try to understand
inflation in the early Universe and accelerated expansion at late times
\citep[see also][]{gegen2016,gegen2018}. This work deals with the Universe
in the large, and is not directly connected with the present paper, which
considers objects of the scale of galaxies or the Solar System.

We note also recent work on the elimination of ghosts in fourth-order
theories \citep{benman1, benman2, benman3}; the presence of ghosts had
previously been a major obstacle to the development of such theories.
These papers deal with the quantum mechanical aspects of fourth-order
theories; the present treatment is purely classical.

\section{Static fields with spherical symmetry}
\label{sec:statssym}

We now specialize further, to static fields with spherical symmetry.
Like Fiedler and Schimming, but apparently independent of them,
\citet{mann04} addressed the problem of the solution of the Bach
equations under these conditions. They found that in addition to the usual
$1/r$ term of the Schwarzschild solution there was a term $\gamma r$.
Mannheim has used this linear potential to obtain a fit to the rotation
curves of galaxies; for a recent paper, see \citet{mann8}. However, the
relevant field equations in the Mannheim model are not the Bach equations,
but the W-E equations, for which a linear potential is not a solution. It
is therefore not clear what use a linear potential can be in such studies,
except as an approximation over a limited range.
\section{Field equations in the linear approximation}
\label{sec:linear}

The most general form for a static metric with spherical symmetry is given
in \citet[equation 8.1.6]{wein2}:
\begin{equation}
\rmd \tau^2 = B(r) \rmd t^2 - A(r) \rmd r^2 - r^2 \left( \rmd \theta^2
+ \sin^2 \theta \,\rmd \phi^2 \right)\,.
\label{eq:sphsymmet}
\end{equation}

For weak fields we write $A(r) = 1 + a(r)$ and $B(r) = 1 + b(r)$,
where $a(r)$ and $b(r)$ are assumed small compared to unity, so that only
terms linear in $a(r)$ and $b(r)$ need be considered.

We will be considering a source such as the Sun, with density $\rho(r)$
and pressure $p(r)$. Within such a source, the pressure terms in
the field equations are much smaller than the density terms and will
normally be omitted. The trace equation then becomes, with primes denoting
differentiation with respect to $r$:
\begin{eqnarray}
\eta R^{\alpha}_{\;\;\alpha} = -\xi T^{\alpha}_{\;\;\alpha}
& = & \xi \rho (r)
\\
-\frac{\eta}{2r^2} \left( 4a - 4r b^{\prime} + 4ra^{\prime}
- 2 r^2 b^{\prime \prime} \right) & = & \xi \rho (r)
\label{eq:trace_2}
\\
2 \left( ra \right)^{\prime} - \left( r^2 b^{\prime} \right)^{\prime}
& = &  -\frac{r^2 \xi \rho (r)}{\eta} \,.
\label{eq:trace_3}
\end{eqnarray}
We assume the density is a smooth, monotonically decreasing function,
so that $a^{\prime}(r)$ and $b^{\prime}(r)$  are both zero at $r=0$.
Then we can integrate out from the origin to $r$ to get
\begin{eqnarray}
2ra - r^2 b^{\prime} & = & -\frac{\xi}{4 \upi \eta}
\int_{0}^{r} 4 \upi u^2 \rho(u)\,\rmd u
\nonumber \\
  & = &  -\frac{\xi}{4 \upi \eta}m_{e} (r) \,,
\label{eq:beqn}
\end{eqnarray}
where $m_e (r)$ is the enclosed mass out to $r$.

The $r,r$ component of the W-E equations gives \footnote{For the
geometrical calculations we have used \textsc{grtensorii} followed by a
\textsc{maple} script to extract the linear terms.}
\begin{eqnarray}
-r^3 b^{\prime \prime \prime} - 2r^2 b^{\prime \prime}
- r^2 a^{\prime \prime} + 2 r b^{\prime} + 2a & & {}
\nonumber \\
{} + 3r^2 \eta \left( -r b^{\prime} + a \right) & = & 0 \,.
\label{eq:rr}
\end{eqnarray}
Before going further, we can check that the Schwarzschild solution is a
possible solution of (\ref{eq:rr}) and the trace equation,
(\ref{eq:trace_2}). This solution is characterised by
$A(r) = 1/B(r) = 1 + \beta/r$, i.e. $a(r) = -b(r) = \beta/r$; we will call
this relation the Schwarzschild condition. \footnote{A few years ago this
writer speculated \citep{prp15} that a second solution of the W-E equations
might also satisfy these Schwarzschild conditions. The present paper
suggests this idea is mistaken.} Substituting these expressions into
(\ref{eq:rr}) and (\ref{eq:trace_2}), we can verify that the equations are
satisfied.

In the limit $\alpha_g \rightarrow \infty$ ($\eta \rightarrow 0$), the Weyl
tensor is everywhere dominant. The trace equations are irrelevant, and
(\ref{eq:rr}) admits the solution
\begin{equation}
a(r) = -b(r) = \gamma r\,,
\end{equation}
the Mannheim linear potential.

These are not, however, the only possibilities, and we will now derive a
different form for $a(r)$ and $b(r)$. The solution we will obtain does not,
of course, guarantee that a corresponding solution exists for the full
nonlinear W-E equations. But it does provide a limiting form, for weak
fields, of such a solution, if it exists.

We will now transform (\ref{eq:rr}) and the trace equation to get a
second-order equation in $a(r)$ only. Differentiating (\ref{eq:trace_2}):
\begin{eqnarray}
-4a^{\prime} -2ra^{\prime \prime}
+ 4r b^{\prime \prime} + r^2 b^{\prime \prime \prime}
+ 2b^{\prime} & = & {}
\nonumber \\
 \frac{2r \xi \rho(r)}{\eta}
+ \frac{r^2 \xi \rho^{\prime} (r)}{\eta} \,. & & {}
\end{eqnarray}
Combining this with (\ref{eq:rr}) we can eliminate
$b^{\prime \prime \prime} (r)$:
\begin{eqnarray}
-3r^2 a^{\prime \prime} - 4r a^{\prime} + 2a
+ 2r^2 b^{\prime \prime} + 4 r b^{\prime} \hspace{0.6in} & & {}
\nonumber \\
{} + 3r^2 \eta \left( -r b^{\prime} + a \right) = 
\frac{2r^2 \xi \rho(r)}{\eta}
+ \frac{r^3 \xi \rho^{\prime} (r)}{\eta} \,. & & {}
\end{eqnarray}

We can now use (\ref{eq:trace_2}) and (\ref{eq:beqn}) to eliminate all
terms involving $b(r)$, to arrive at
\begin{equation}
a^{\prime \prime} + \left( -\frac{2}{r^2} + \eta \right) a = 
- \frac{\xi}{4 \upi r} m_e (r)
- \frac{r \xi \rho^{\prime} (r)}{3 \eta} \equiv -H \,.
\label{eq:2ord}
\end{equation}

Associated with this equation is the homogeneous equation
\begin{equation}
a^{\prime \prime} + \left( -\frac{2}{r^2} + \eta \right) a = 0 \,.
\label{eq:2ordhom}
\end{equation}

The appearance of the derivative of the density on the right side of
(\ref{eq:2ord}) reminds us that although this equation is second order, it
originates in the third-order equation (\ref{eq:rr}), and therefore will
probe more intimately into the density distribution than we are familiar
with in conventional general relativity. Indeed, as we will see in sections
\ref{sec:ag1} -- \ref{sec:alincom}, it is precisely this term, through the
associated integrals ${\mathcal H}_2 $ and ${\mathcal J}_2 $, that causes
us the most trouble in practice.

At this point we choose $\alpha_g < 0$, and therefore $\eta < 0$.
This will ensure that we deal with modified Bessel functions, which have
a particularly simple form. Also, because in this paper we are looking for
solutions analogous to Mannheim's linear potential, we will assume,
tentatively, that the length $r_0 = 1/k$ is of galactic scale, intermediate
between the scale of the SS and truly cosmological scales.

We will develop the solution of (\ref{eq:2ord}) as an integral over the
source density, using a Green function constructed from the related
homogeneous equation, which now can be written
\begin{equation}
a^{\prime \prime} - \left( \frac{\nu^2 - 1/4}{r^2} + k^2 \right) a
= 0 \,,
\label{eq:2ordhom2}
\end{equation}
with $\nu = 3/2$ and $k^2 = -\eta > 0$. We use the notation of
\citet{absteg} (AS in what follows). For the solution
of (\ref{eq:2ordhom2}), AS, equation 9.1.49, and the paragraph preceding 
9.6.41, gives $a(r) = r^{1/2} {\mathcal L}_{3/2} (kr)$, where
${\mathcal L}_{\nu}$ stands for $I_{\nu}$ or $K_{\nu}$. We can express
these solutions in terms of spherical Bessel functions. Setting $kr = z$,
and using AS, equations 10.2.13 and 10.2.17, we define:
\begin{eqnarray}
a_I (z) & \equiv & \frac{2}{\upi} z^{1/2} I_{3/2} (z)
\nonumber \\
  & = & \left( -\frac{\sinh z}{z} + \cosh z \right)
\label{eq:aimsb}
\\
a_K (z) & \equiv & \frac{2}{\upi} z^{1/2} K_{3/2} (z)
\nonumber \\
  & = & \left( 1 + \frac{1}{z}  \right) \rme^{-z}
\label{eq:akmsb}
\\
W[a_I (z), a_K (z)] & \equiv & a_I (z) \frac{ da_{K} (z)}{dz}
- \frac{ da_{I} (z)}{dz} a_K (z)
\nonumber \\
  & = & -1
\hspace{0.25in}\mbox{(Wronskian)}\,.
\label{eq:wronsk}
\end{eqnarray}

Associated with these metric functions are $b_I (z)$ and $b_K (z)$, obtained
by integrating the trace equation (\ref{eq:beqn}):
\begin{eqnarray}
b_I (z) & = & \frac{2 \sinh z}{z}
\label{eq:bimsb}
\\
b_K (z) & = & -\frac{2 \rme^{-z}}{z}
\label{eq:bkmsb}
\end{eqnarray}

If we can find a Particular Integral (PI) of our equation (\ref{eq:2ord}),
the general solution is the PI plus a Complementary Function (CF) that is a
solution of the homogeneous equation (\ref{eq:2ordhom}). This construction
is useful only if a PI can actually be found, but this turns out to be the
case for our problem.

\section{The Green function }
\label{sec:green}

We will consider two basic forms for our Green function, $G_1 (y,z)$ and
$G_2 (y,z)$, where $y$ and $z$ are both positive. $y$ refers to the source
point, and $z$ to the field point. We will use a range of radii from
$r_{\rm min}$ to $r_{\rm max}$, and anticipate that we may be able to let
$r_{\rm min}$ tend to zero, if all metric functions are regular at the
origin. But we will have to be cautious about letting $r_{\rm max}$ tend to
infinity if we have to deal with a potential that rises indefinitely, like
Mannheim's linear potential.

For a compact source such as the Sun, our main concern in this paper,
$G_1 $ has the form \citep[see][]{arfken2}:
\renewcommand{\arraystretch}{2.0}
\begin{eqnarray}
G_1 (y,z) & = & \left\{ \begin{array}{ll}
{\displaystyle \frac{ a_I (y) a_K (z)}{k}},
\hspace{0.25in}r_{\rm min} \le y < z\,, \\
{\displaystyle \frac{ a_K (y) a_I (z)}{k}},
\hspace{0.25in}z < y < r_{\rm max}\,.
\end{array} \right.
\end{eqnarray}
\renewcommand{\arraystretch}{1.0}

$a(r)$ is then given as the integral:
\begin{equation}
a(r) = \int_{r_{\rm min}}^{r_{\rm max}} G_1 (kr,kt) H(t)\,\rmd t \,.
\end{equation}

Our second Green function is
\renewcommand{\arraystretch}{2.0}
\begin{eqnarray}
G_2 (y,z) & = & \left\{ \begin{array}{ll}
{\displaystyle -\frac{ a_K (y) a_I (z)}{k}},
\hspace{0.25in}r_{\rm min} \le y < z\,, \\
{\displaystyle -\frac{ a_I (y) a_K (z)}{k}},
\hspace{0.25in}z < y < r_{\rm max}\,.
\end{array} \right.
\end{eqnarray}
\renewcommand{\arraystretch}{1.0}
with $a(r)$ now given by
\begin{equation}
a(r) = \int_{r_{\rm min}}^{r_{\rm max}} G_2 (kr,kt) H(t)\,\rmd t \,.
\end{equation}

More generally, we can consider a combination of $G_1 $ and $G_2 $:
\begin{equation}
G(y,z) = {\rm P} G_1 (y,z) + ( 1 - {\rm P}) G_2 (y,z)\,,
\end{equation}
where ${\rm P}$ can be chosen to satisfy the constraints of our problem. The
condition that the metric functions be regular at the origin is easily met,
and the conditions at $r_{\rm max}$ are what we are trying to discover; the
important constraint is that in the SS the metric functions satisfy what
we will call the Schwarzschild condition, to be discussed more
fully later, in section \ref{sec:schw}.

\section{The source}
\label{sec:source}

In this paper we will be mainly concerned with the SS, so our source is the
Sun. However, when dealing with any compact system with a long-range
potential we have to worry about contributions from distant matter. Let us
call this the ``embedding problem''. For galaxies, Mannheim has proposed a
solution \citep[see][section 9.3]{mann6}, and there is a long tradition
stemming from the paper of \citet{einstra} (the ``Einstein vacuole'').

But we have to devise a reasonable embedding for the SS. We should not
simply assume a compact mass at the center (the Sun) surrounded by vacuum.
The SS is embedded in a galaxy (the Milky Way), with a mean background
density $\rho_b$. If the SS had not yet formed, there would be negligible
gravitational field at its location. So it is reasonable to suppose, at
least in the linear approximation, that what best characterises the source
is the {\em difference} between $\rho_b $ and the actual density in the SS.
For simplicity let us suppose the SS formed by the collapse of material
enclosed within a sphere of radius $r_v$ ($v$ for vacuole), while material
at a larger radius is unaffected. $r_v $, of course, is much larger than
$r_s $, the radius of the Sun. Then we have
\begin{equation}
\msun = 4\upi \int_{0}^{r_s} r^2 \rho (r)\,\rmd r = \frac{4 \upi \rho_b
r_{v}^{3}}{3} \,.
\end{equation}
In the range $r_s < r < r_v $ we must use a constant {\em negative} density,
$-\rho_b $.

\section{Numbers}
\label{sec:numbers}

We give here some numbers for the SS and other objects; this is mainly for
orientation, since we will not make much actual use of the numbers:
\begin{eqnarray}
\mbox{background density, }\rho_b &= & 1.5 \times 10^{-23}\,
{\rm g \cdot cm}^{-3}
\nonumber \\
\mbox{mean solar density, }\rho_s & = & 1.4\,{\rm g \cdot cm}^{-3}
\nonumber \\
\mbox{radius of the Sun, }r_s & = & 7 \times 10^{10}\,{\rm cm}
\nonumber \\
\mbox{radius of Neptune's orbit, }r_N & = & 4.5 \times 10^{14}\,{\rm cm}
\nonumber \\
\mbox{radius of the vacuole, }r_v & = & 3.6 \times 10^{18}\,{\rm cm}
\nonumber \\
\mbox{size of the Galaxy, }r_0 & = & 10^{23}\,{\rm cm}\,.
\end{eqnarray}

We will take $r_{\rm max}$ to be $100 r_0 = 10^{25}\,{\rm cm}$. The two
regions of interest us are

${\mathcal R}_1$. $r_s < r < r_v$; this region includes the SS.

${\mathcal R}_2$. $r_v < r < r_{\rm max}$; this region is outside all
sources.

For the sake of completeness, we will retain terms involving $\rho_b $. We
note here, however, that in the SS the effect of these terms is very small;
for example, at $r = r_N $, the related acceleration is about seven orders
of magnitude smaller than that associated with the Pioneer anomaly
\citep{tury}, which is itself near the limit of detectability.

\section{The metric in region $\bmath{\mathcal R}_1$, general
considerations}
\label{sec:metA}

In this region, we have
\begin{equation}
m_e (r) = \msun - \frac{4 \upi \rho_b r^3}{3}
\hspace{0.25in}{\rm and}\hspace{0.25in}\rho^{\prime} = 0\,,
\end{equation}
and a suitable PI for (\ref{eq:2ord}) is
\begin{equation}
a_{\rm PI} (r) = \frac{\xi \msun}{4 \upi k^2 r}
- \frac{\xi \rho_b r^2}{3 k^2} \,.
\label{eq:aPI}
\end{equation}
Since $r_0$ is assumed to be of galactic scale, $kr$ is much less than
unity for $r$ within the SS.

Associated with this metric function is $b_{\rm PI} (r)$, obtained
by integrating the trace equation (\ref{eq:beqn}):
\begin{eqnarray}
b_{\rm PI} (r) & = & -\frac{\xi \msun}{4 \upi k^2 r}
- \frac{\xi \rho_b r^2 }{6 k^2 } + {\mathcal E}\,,
\label{eq:bPI}
\end{eqnarray}
where ${\mathcal E}$ is an integration constant chosen to ensure
$b_{\rm PI} (r_v ) = 0$:
\begin{equation}
{\mathcal E} = \frac{\xi \msun}{4 \upi k^2 r_v }
+ \frac{\xi \rho_b r_{v}^{2}}{6 k^2 } \,.
\label{eq:cale}
\end{equation}

In equations (\ref{eq:aPI}) and (\ref{eq:bPI}) the most important terms, for
$r$ within the SS, are the first ones. If we neglect the others, we see that
these first terms give us just the Schwarzschild solution.

We will write the CF as
\begin{equation}
{\rm CF} = {\mathcal B}_I a_I (kr) + {\mathcal B}_K a_K (kr) \,.
\end{equation}

We can verify that our original third-order equation, (\ref{eq:rr}),
is solved not only by the PI pair, $a_{\rm PI} (r)$, $b_{\rm PI} (r)$, but
also by the CF pairs $a_I (r)$, $b_I (r)$ and $a_K (r)$, $b_K (r)$.

Our Green function will determine the constants ${\mathcal B}_I$ and
${\mathcal B}_K$. It will be constructed to satisfy certain conditions, the
most important of which is discussed in section \ref{sec:schw}.

\section{The metric in region $\bmath{\mathcal R}_2$, general
considerations}
\label{sec:metB}

\begin{figure}
\centering
\includegraphics[scale=0.27,angle=-90]{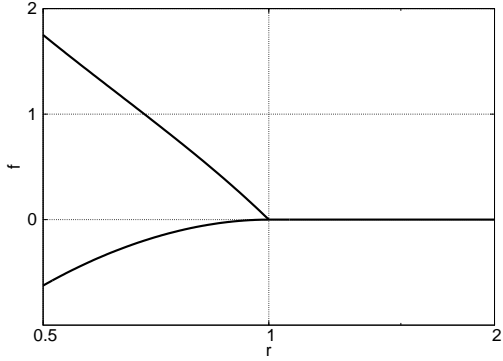}
\caption{\label{fig:pi}This figure illustrates the behavior of the PI's,
using (upper curve) $a_{\rm PI} = 1/r - r^2$ for $r< 1$, and
$a_{\rm PI} = 0$ for $r> 1$, and (lower curve)
$b_{\rm PI} = -1/r - r^2 /2 + 3/2$ for $r< 1$, and $b_{\rm PI} = 0$
for $r> 1$. }
\end{figure}

In this region, $m_e (r) = \rho^{\prime} = 0$, so the PI is also zero. The
general behaviour of the PI across the boundary at $r = r_v $ is shown in
figure \ref{fig:pi}. The graph, with its discontinuity in slope, is
reminiscent of the potential in problems of electrostatics.

The CF, however, being the solution of a homogeneous equation, is
independent of sources, and will show no discontinuity of any kind across
the boundary.

\section{The Schwarzschild condition}
\label{sec:schw}

Observations tell us that to high accuracy the metric in the SS is of
Schwarzschild form, according to which $A(r) = 1/B(r)$, or, for weak fields,
$a(r) = -b(r)$. This we will call the Schwarzschild condition.
A solution that is pure PI represents the usual
Schwarzschild solution, so we have to ask what CF we can add and still
preserve the Schwarzschild condition.

Making ${\mathcal B}_I$ non-zero is allowed, providing it is not too large,
because $a_I (z) \approx z^2 /3$ for small $z$. (See section
\ref{sec:alincom} for a discussion of ${\mathcal B}_I$). But with
${\mathcal B}_K$ we have to be careful, because $a_K (z) \approx 1/z$,
similar to the PI. So let us examine a solution of the form
\begin{eqnarray}
a(r) &= & {\rm PI} + {\mathcal B}_K a_K (kr)
\nonumber \\
  & = & \frac{\xi \msun}{4 \upi k^2 r} - \frac{\xi \rho_b r^2 }{3 k^2}
+ {\mathcal B}_K \left( 1 + \frac{1}{kr} \right) \rme^{-kr}
\nonumber \\
  & \approx & \frac{\xi \msun}{4 \upi k^2 r}
+ {\mathcal B}_K \left( \frac{1}{kr} \right)
\end{eqnarray}
for small $kr$. We can derive $b(r)$ from (\ref{eq:beqn}). For points in the
SS;
\begin{eqnarray}
b^{\prime} (r) & = & \frac{2a}{r} + \frac{\xi \msun}{4 \upi \eta r^2}
\nonumber \\
  & \approx & \frac{\xi \msun}{2 \upi k^2 r^2 }
+ {\mathcal B}_K \left(\frac{2}{kr^2} \right)
- \frac{\xi \msun }{4 \upi k^2 r^2 }
\nonumber \\
  & \approx & \frac{\xi \msun}{4 \upi k^2 r^2 } + {\mathcal B}_K
\left(\frac{2}{kr^2 }\right)
\nonumber \\
b(r) & = & -\frac{\xi \msun}{4 \upi k^2 r} - {\mathcal B}_K \left(\frac{2}{kr}
\right) \,.
\end{eqnarray}
We note that the terms derived from the PI satisfy the Schwarzschild
condition, as expected, but the terms involving ${\mathcal B}_K$ do not.
Our Green function must therefore be constructed to minimise
$\left|{\mathcal B}_K \right|$.

\section{The metric in region ${\mathcal R}_1$, using $\bmath{G_1}$}
\label{sec:ag1}

We will define our Green function over the region $r_{\rm min} < r < r_v$,
i.e. we set $r_{\rm max} = r_v$. In practice, we will be able to let
$r_{\rm min} \rightarrow 0$ in the end, so this region is larger than
region ${\mathcal R}_1$, because it includes $0 < r < r_s$, the interior of
the Sun. Note that the range of $G_1 $ does not include the step in density
at $r_v$; let us use the notation $r_{v-}$ to emphasise this point.

In the body of this paper we assume the Weyl limit within the SS, $kr \ll 1$.
But in appendix \ref{sec:check} we check the Green function $G_1 $ using
the Einstein limit, $kr \gg 1$, where we know the solution.

Our metric function in region ${\mathcal R}_1$, $a(r)$, can be written as
four integrals:
\begin{eqnarray}
a(r) & = & a_< (r) + a_> (r) \,,
\end{eqnarray}
where
\begin{eqnarray}
a_< (r) & = & \frac{a_K (kr)}{k} \left[{\mathcal H}_1 (r) + {\mathcal H}_2 (r)
\right]
\nonumber \\
{\mathcal H}_1 (r) & = & \int_{r_{\rm min}}^{r} a_I (kt) 
\left[\frac{\xi}{4 \upi t} m_e (t) \right]\,\rmd t
\label{eq:h1_msb}
\\
{\mathcal H}_2 (r) & = & \int_{r_{\rm min}}^{r} a_I (kt) 
\left[\frac{t \xi \rho^{\,!} (t)}{3 \eta} \right] \,\rmd t\,,
\label{eq:h2_msb}
\end{eqnarray}
and
\begin{eqnarray}
a_> (r) & = & \frac{a_I (kr)}{k} \left[{\mathcal H}_3 (r) + {\mathcal H}_4 (r)
\right]
\nonumber \\
{\mathcal H}_3 (r) & = & \int_{r}^{r_{v-}} a_K (kt) 
\left[\frac{\xi}{4 \upi t} m_e (t) \right]\,\rmd t
\label{eq:h3_msb}
\\
{\mathcal H}_4 (r) & = & \int_{r}^{r_{v-}} a_K (kt) 
\left[\frac{t \xi \rho^{\,!} (t)}{3 \eta} \right] \,\rmd t\,.
\label{eq:h4_msb}
\end{eqnarray}

The last of our four integrals, ${\mathcal H}_4 $, is clearly zero.
The first, ${\mathcal H}_1 $, will be split by writing
\begin{eqnarray}
{\mathcal H}_1 & = &
\frac{\xi }{4 \upi } \int_{kr_s}^{kr} \frac{a_I (z)}{z}
\left( \msun - \frac{4 \upi \rho_b z^3 }{3 k^3 } \right) \, \rmd z
+ {\mathcal H}_{\rm sun}\,,
\label{eq:H1ref}
\end{eqnarray}
where
\begin{eqnarray}
{\mathcal H}_{\rm sun} & \approx & 
\frac{\xi }{4 \upi } \int_{r_{\rm min}}^{r_s} \frac{a_I (kt)}{t}
m_e (t) \, \rmd t
\end{eqnarray}
depends on the density distribution within the Sun, as modelled in the
Appendix.

The integrals in (\ref{eq:H1ref}) are straightforward:
\begin{eqnarray}
{\mathcal H}_1 & = & \frac{\xi}{4 \upi} \left. \overline{\mathcal H}_1 (z)
\right|_{z=kr_s}^{z=kr} + {\mathcal H}_{\rm sun}\,,
\hspace{0.25in}{\rm where}
\nonumber \\
\overline{\mathcal H}_1 (z) & = & \msun \frac{\sinh z}{z}
- \frac{4 \upi \rho_b }{3 k^3} \left( z^2 \sinh z \right.
\nonumber \\
  & & \left. {} - 3z \cosh z + 3 \sinh z \right) \,.
\end{eqnarray}

It will be convenient in what follows to divide ${\mathcal H}_1 $ further,
according to the limits on $\overline{\mathcal H}_1 (z)$:
\begin{eqnarray}
{\mathcal H}_1 & = & {\mathcal H}_5 + {\mathcal H}_6
+ {\mathcal H}_{\rm sun}\,, \hspace{0.25in}
{\rm where}
\nonumber \\
{\mathcal H}_5 & = & -\frac{\xi }{4 \upi} \overline{\mathcal H}_1 (kr_s)
\label{eq:H5ref}
\\
{\mathcal H}_6 & = & \frac{\xi }{4 \upi} \overline{\mathcal H}_1 (kr) \,.
\label{eq:H6ref}
\end{eqnarray}

For ${\mathcal H}_2 $ we set $a_I (kt)$ equal to its limiting value for small
$kt$, namely $k^2 t^2 /3$:
\begin{eqnarray}
{\mathcal H}_2 & = & \frac{\xi k^2 }{9 \eta} 
\int_{r_{\rm min}}^{r} t^3 \rho^{\prime} \,\rmd t
\nonumber \\
  & = & - \frac{\xi}{9 } 
\left[\left. t^3 \rho (t) \right|_{r_{\rm min}}^{r}
- \int_{r_{\rm min}}^{r} 3t^2 \rho (t) \,\rmd t \right]
\nonumber \\
  & \stackrel[r_{\rm min} \rightarrow 0]{}{\longrightarrow} &
\frac{\xi \msun }{12 \upi }  \,.
\end{eqnarray}

Finally, ${\mathcal H}_3 $ evaluates to:
\begin{equation}
{\mathcal H}_3 = \frac{\xi }{4 \upi}
\int_{kr}^{kr_{v-}} \frac{a_K (z)}{z}
 \left( \msun - \frac{4 \upi \rho_b z^3}{3k^3} \right)
\,\rmd z\,.
\label{eq:H3ref}
\end{equation}
As with ${\mathcal H}_1$, the integrals in (\ref{eq:H3ref}) are
straightforward, and lead to
\begin{eqnarray}
{\mathcal H}_3 & = & \frac{\xi }{4 \upi}
\left. \overline{\mathcal H}_3 (z) \right|_{z=kr}^{z = kr_{v-}}
\nonumber \\
\overline{\mathcal H}_3 (z) & = & -\msun\frac{\rme^{-z}}{z}
+ \frac{4\upi \rho_b}{3k^3 }
\left(z^2 + 3z + 3 \right) \rme^{-z}\,.
\end{eqnarray}

We divide ${\mathcal H}_3$ according to the limits on the integral:
\begin{eqnarray}
{\mathcal H}_3 & = & {\mathcal H}_7 + {\mathcal H}_8\,,
\hspace{0.25in}{\rm where}
\nonumber \\
{\mathcal H}_7 & = & -\frac{\xi}{4\upi} \overline{\mathcal H}_3 (z=kr)
\nonumber \\
{\mathcal H}_8 & = & \frac{\xi}{4\upi} \overline{\mathcal H}_3 (z=kr_{v-})\,.
\end{eqnarray}
We note that ${\mathcal H}_6$ and ${\mathcal H}_7$ are functions of $r$,
and together produce the PI of (\ref{eq:aPI}).

\begin{figure}
\centering
\includegraphics[scale=0.27,angle=-90]{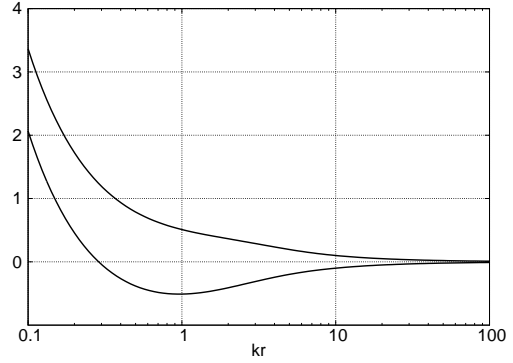}
\caption{\label{fig:ab} Upper (lower) curve: $a$ ($b$) as a function of
$kr$. The vertical scale is arbitrary. $a(r)$ is taken from (\ref{eq:aapp}),
and $b(r)$ from (\ref{eq:bapp}). }
\end{figure}

${\mathcal H}_2$, ${\mathcal H}_5$, ${\mathcal H}_8$
and ${\mathcal H}_{\rm sun}$ are constants, so
we will write the CF for this Green function as
\begin{eqnarray}
{\rm CF} & = & {\mathcal C}_I a_I (kr) + {\mathcal C}_K a_K (kr) \,,
\hspace{0.25in}{\rm where}
\nonumber \\
{\mathcal C}_I & = & \frac{1}{k} {\mathcal H}_8
\nonumber \\
{\mathcal C}_K & = & \frac{1}{k} \left( {\mathcal H}_2
+ {\mathcal H}_5 + {\mathcal H}_{\rm sun} \right)\,.
\end{eqnarray}
We can omit the term ${\mathcal H}_{\rm sun}$, which is smaller by a
factor of order $k^2 r_{s}^2$ than the other two terms (see
appendix \ref{sec:appb} ). This gives:
\begin{eqnarray}
{\mathcal C}_K & = & \frac{1}{k} \left({\mathcal H}_2 + {\mathcal H}_5 \right)
\; = \; -\frac{\xi \msun}{6 \upi k}\,.
\end{eqnarray}
For the metric function $a(r)$ in the SS we can omit the small contribution
from ${\mathcal C}_I$ and the term in $\rho_b$ in the PI:
\begin{eqnarray}
a(r) & = & \frac{\xi \msun}{4 \upi k^2 r}
- \frac{\xi \msun}{6 \upi k } \left(1 + \frac{1}{kr} \right) \rme^{-kr}
\label{eq:aapp}
\\
  & \approx & \frac{\xi \msun}{12 \upi k^2 r}\,,\;\mbox{for $kr \ll 1$.}
\end{eqnarray}
If we ignore terms in $\rho_b$ for the moment, and simply use
(\ref{eq:aapp}) in (\ref{eq:beqn}), we can integrate to get
\begin{equation}
b(r) = -\frac{\xi \msun}{4 \upi k^2 r}
+ \frac{\xi \msun}{3 \upi k^2 r} \rme^{-kr}\,,
\label{eq:bapp}
\end{equation}
which reduces, in the SS, to
\begin{equation}
b(r) \approx \frac{\xi \msun}{12 \upi k^2 r} \,.
\end{equation}
$a(r)$, from (\ref{eq:aapp}), and $b(r)$, from (\ref{eq:bapp}), are
plotted in figure \ref{fig:ab}.

${\mathcal C}_K$ is non-zero, so $G_1 $ by itself is not an acceptable Green
function; it does not lead to the Schwarzschild condition in the SS,
$b(r) = -a(r)$. Instead we have $b(r) \approx a(r)$.

\section{The metric in region $\bmath{\mathcal R}_1$, using $\bmath{G_2}$}
\label{sec:ag2}

Here we recapitulate the previous section, with appropriate changes in the
integrals.

Our metric function in region ${\mathcal R}_1$, $a(r)$, can be written as
four integrals:
\begin{eqnarray}
a(r) & = & a_< (r) + a_> (r) \,,
\end{eqnarray}
where
\begin{eqnarray}
a_< (r) & = & -\frac{a_I (kr)}{k} \left[{\mathcal J}_1 (r)
+ {\mathcal J}_2 (r) \right]
\nonumber \\
{\mathcal J}_1 (r) & = & \int_{r_{\rm min}}^{r} a_K (kt) 
\left[\frac{\xi}{4 \upi t} m_e (t) \right]\,\rmd t
\label{eq:j1_msb}
\\
{\mathcal J}_2 (r) & = & \int_{r_{\rm min}}^{r} a_K (kt) 
\left[\frac{t \xi \rho^{\,!} (t)}{3 \eta} \right] \,\rmd t\,,
\label{eq:j2_msb}
\end{eqnarray}
and
\begin{eqnarray}
a_> (r) & = & -\frac{a_K (kr)}{k} \left[{\mathcal J}_3 (r)
+ {\mathcal J}_4 (r) \right]
\nonumber \\
{\mathcal J}_3 (r) & = & \int_{r}^{r_{v-}} a_I (kt) 
\left[\frac{\xi}{4 \upi t} m_e (t) \right]\,\rmd t
\label{eq:j3_msb}
\\
{\mathcal J}_4 (r) & = & \int_{r}^{r_{v-}} a_I (kt) 
\left[\frac{t \xi \rho^{\,!} (t)}{3 \eta} \right] \,\rmd t\,.
\label{eq:j4_msb}
\end{eqnarray}

The last of our four integrals, ${\mathcal J}_4 $, is clearly zero. The
first, ${\mathcal J}_1 $, will be split by writing:
\begin{equation}
{\mathcal J}_1 = \frac{\xi }{4 \upi } \int_{kr_s}^{kr} \frac{a_K (z)}{z}
\left( \msun - \frac{4 \upi \rho_b z^3 }{3 k^3 } \right) \, \rmd z
+ {\mathcal J}_{\rm sun}\,.
\label{eq:J1ref}
\end{equation}
where
\begin{eqnarray}
{\mathcal J}_{\rm sun} & \approx & 
\frac{\xi }{4 \upi } \int_{r_{\rm min}}^{r_s} \frac{a_K (kt)}{t}
m_e (t) \, \rmd t
\end{eqnarray}
depends on the density distribution within the Sun, as modelled in the
Appendix.

The integrals in (\ref{eq:J1ref}) are straightforward, and result in
\begin{eqnarray}
{\mathcal J}_1 & = & \frac{\xi}{4 \upi} \left.\overline{\mathcal J}_1 (z)
\right|_{z = kr_s}^{z = kr} + {\mathcal J}_{\rm sun}
\nonumber \\
\overline{\mathcal J}_1 (z) & = & -\msun\frac{\rme^{-z}}{z}
+ \frac{4\upi \rho_b}{3k^3 }
\left(z^2 + 3z + 3 \right) \rme^{-z} \,.
\end{eqnarray}

It will be convenient in what follows to divide ${\mathcal J}_1 $ further,
according to the limits on $\overline{\mathcal J}_1 (z)$:
\begin{eqnarray}
{\mathcal J}_1 & = & {\mathcal J}_5 + {\mathcal J}_6
+ {\mathcal J}_{\rm sun}\,, \hspace{0.25in}
{\rm where}
\nonumber \\
{\mathcal J}_5 & = & -\frac{\xi }{4 \upi} \overline{\mathcal J}_1 (kr_s)
\label{eq:J5ref}
\\
{\mathcal J}_6 & = & \frac{\xi }{4 \upi} \overline{\mathcal J}_1 (kr) \,.
\label{eq:J6ref}
\end{eqnarray}

For ${\mathcal J}_2 $ we set $a_K (kt)$ equal to its limiting value for small
$kt$, namely $1/(kt)$:
\begin{eqnarray}
{\mathcal J}_2 & = & \frac{\xi }{3 \eta } 
\int_{r_{\rm min}}^{r} \left(\frac{1}{kt} \right) t \rho^{\prime} \,\rmd t
\nonumber \\
  & \stackrel[r_{\rm min} \rightarrow 0]{}{\longrightarrow} &
- \frac{\xi}{3 \eta k} \rho(0) \,.
\end{eqnarray}

Finally, ${\mathcal J}_3 $ evaluates to:
\begin{equation}
{\mathcal J}_3 = \frac{\xi }{4 \upi}
\int_{kr}^{kr_{v-}} \frac{a_I (z)}{z}
 \left( \msun - \frac{4 \upi \rho_b z^3}{3k^3} \right)
\,\rmd z\,.
\label{eq:J3ref}
\end{equation}
As with ${\mathcal J}_1$, the integrals in (\ref{eq:H3ref}) are
straightforward, and lead to
\begin{eqnarray}
{\mathcal J}_3 & = & \frac{\xi }{4 \upi}
\left. \overline{\mathcal J}_3 (z) \right|_{z=kr}^{z = kr_{v-}}
\nonumber \\
\overline{\mathcal J}_3 (z) & = & \msun \frac{\sinh z}{z}
- \frac{4 \upi \rho_b }{3 k^3} \left( z^2 \sinh z \right.
\nonumber \\
  & & \left. {} - 3z \cosh z + 3 \sinh z \right)\,.
\end{eqnarray}

We divide ${\mathcal J}_3$ according to the limits on the integral:
\begin{eqnarray}
{\mathcal J}_3 & = & {\mathcal J}_7 + {\mathcal J}_8\,,
\hspace{0.25in}{\rm where}
\nonumber \\
{\mathcal J}_7 & = & -\frac{\xi}{4\upi} \overline{\mathcal J}_3 (z=kr)
\nonumber \\
{\mathcal J}_8 & = & \frac{\xi}{4\upi} \overline{\mathcal J}_3 (z=kr_{v-})
\;\approx\; \frac{\xi \msun}{4 \upi}\,.
\end{eqnarray}
We note that ${\mathcal J}_6$ and ${\mathcal J}_7$ are functions of $r$,
and together produce the PI of (\ref{eq:aPI}).

${\mathcal J}_2$, ${\mathcal J}_5$, ${\mathcal J}_8$
and ${\mathcal J}_{\rm sun}$ are constants, and (as for
${\mathcal H}_{\rm sun}$) ${\mathcal J}_{\rm sun}$ is negligible in
comparison to the others. So we will write the CF for this Green function as
\begin{eqnarray}
{\rm CF} & = & {\mathcal D}_I a_I (kr) + {\mathcal D}_K a_K (kr) \,,
\hspace{0.25in}{\rm where}
\nonumber \\
{\mathcal D}_I & = & -\frac{1}{k} \left( {\mathcal J}_2
+ {\mathcal J}_5 \right)
\nonumber \\
{\mathcal D}_K & = & -\frac{1}{k} {\mathcal J}_8
\;\approx\; -\frac{\xi \msun}{4 \upi k}\,.
\end{eqnarray}
${\mathcal D}_K$ is non-zero, so $G_2 $ by itself is not an acceptable Green
function; it does not lead to the Schwarzschild condition in the SS.

\section{The metric in region $\bmath{\mathcal R}_1$, using a linear
combination of $\bmath{G_1}$ and $\bmath{G_2}$}
\label{sec:alincom}

Both ${\mathcal C}_K$ and ${\mathcal D}_K$ are non-zero, so we have to
construct a linear combination of $G_1$ and $G_2$,
$G = {\rm P}G_1 + (1-{\rm P})G_2$. This preserves the PI, and for the
coefficient ${\mathcal B}_K$ in the CF we get
\begin{eqnarray}
{\mathcal B}_K & = & {\rm P}{\mathcal C}_K + (1-{\rm P}){\mathcal D}_K
\nonumber \\
  & = & -\frac{\xi \msun}{12 \upi k} \left[2{\rm P} + 3(1 - {\rm P} )\right]
\nonumber \\
  & = & 0 \hspace{0.25in}{\rm if} \hspace{0.25in}{\rm P} = 3\,,
\end{eqnarray}
so that the coefficient ${\mathcal B}_I$ is given by
\begin{eqnarray}
{\mathcal B}_I & = & 3 {\mathcal C}_I - 2{\mathcal D}_I
\nonumber \\
  & = & \frac{3}{k} {\mathcal H}_8
+ \frac{2}{k} \left( {\mathcal J}_2 + {\mathcal J}_5 \right)
\nonumber \\
  & = & \frac{3 \xi}{4 \upi k} \left(
-\frac{\msun}{k r_{v-}} + \frac{4 \upi \rho_b }{k^3 } \right)
\nonumber \\
  &  & {} + \frac{2}{k} \left[ -\frac{\xi \rho (0) }{3 \eta k }
- \frac{\xi}{4 \upi} \left( -\frac{\msun}{k r_s}
+ \frac{4 \upi \rho_b }{k^3 } \right) \right]\,.
\end{eqnarray}
On the right side of this equation we need retain only the dominant term,
the one derived from ${\mathcal J}_2$:
\begin{eqnarray}
{\mathcal B}_I & = & - \frac{2 \xi \rho (0) }{3 \eta k^2 }\,,
\end{eqnarray}
so our CF is
\begin{eqnarray}
{\rm CF} & = & - \frac{2 \xi \rho (0) }{3 \eta k^2 } a_I (kr)
\nonumber \\
  & \approx & - \frac{2 \xi \rho (0) r^2}{9 \eta }
\end{eqnarray}
within the SS. Combining this with the PI, we obtain for the metric function
$a(r)$ within the SS;
\begin{eqnarray}
a(r) & \approx & - \frac{\xi \msun}{4 \upi \eta r}
- \frac{2 \xi \rho (0) r^2}{9 \eta }\,,
\label{eq:ass}
\end{eqnarray}
where we have omitted the small term in $\rho_b $. 

The second term on the right side of (\ref{eq:ass}) will produce the
long-range potential analogous to the linear potential of Mannheim. But we
can see already that it is unacceptably large. With our assumption that the
density is a maximum at the origin, and is monotonically decreasing, 
$\rho(0) r_{s}^{3} $ is of order $\msun$. So this second term is of order
$\xi \msun / (\eta r_{s}^{3} )$, and the magnitude of the second term
exceeds that of the first, Newtonian, term by a factor of about
$(r/r_s)^3$. Taking $r$ to be the radius of the Earth's orbit, this is
$1500^3 \approx 3 \times 10^9$.

\citet{mann6} takes the coefficients of the Schwarzschild and the linear
potentials to be independent, and determined from observation. Our Green
function approach, on the other hand, shows these coefficients are
connected, so that once we know the Schwarzschild coefficient (Mannheim's
$\beta^{*}$) we know not only the PI but also the CF, depending on our
choice of Green function. This CF, moreover, turns out to be in conflict
with observations of the Solar System.

\section{\label{sec:newapp}Can the W-E equations represent reality?}

At the beginning of this paper we pointed out that a linear potential, used
by \citet{mann8} in their studies of galactic rotation, was not a solution
of the W-E equations, which are the relevant field equations for Mannheim's
model. We then began to search for a solution of the \linebreak
W-E equations that
might include a term that approximates a linear potential. Specialising to
weak, static fields with spherical symmetry, and choosing the critical
length $r_0 $ to be of galactic scale, we used a Green function approach to
construct the solution of the linearised W-E equations for a compact source
such as the Sun (assumed to have a density that is a monotonically
decreasing function of radius). This solution is in conflict with
observations of the Solar System; either it does not have the required
Schwarzschild form, or it has an unacceptably large contribution from the
long-range function $a_I (kr)$.

This does not mean, however, that the W-E equations are useless.
We should simply discard our initial assumption, that $r_0 = 1/k$ is of
galactic scale. Indeed, it would be surprising if a SBT resulted in so
large a value of $r_0 $. More likely would seem to be a value of order of
the size of elementary particles, $10^{-15}\,{\rm m}$ or less. In this
case the Einstein equations would be adequate at all scales accessible to
experiment, and the Schwarzschild solution would be appropriate for the
Solar System. We have seen in appendix \ref{sec:check} that our Green
function correctly identifies the Schwarzschild solution in this limit.

The W-E equations could still have important theoretical applications,
however, because at the highest energies we expect the SBT to be reversed,
so that we recover the original conformal form in which all coupling
constants are dimensionless. The theory is then potentially renormalisable.

\section*{acknowledgments}
This paper has benefitted from suggestions made by an anonymous reviewer.
The author wishes to acknowledge support provided by Washington University
for a retired faculty member.


\appendix

\section{Checking the Green function in the Einstein limit}
\label{sec:check}

In the Einstein limit, $|\xi|$ and $|\eta|$ both become large, but in such a
way that their ratio stays the same. For $r$ in the SS, $kr \gg 1$.
Our metric function, $a(r)$, can be written in terms of the Green function
$G_1$:
\begin{eqnarray}
a(r) & = & a_< (r) + a_> (r)
\nonumber \\
a_< (r) & = & \frac{a_K (kr)}{k} \int_{r_{\rm min}}^{r} a_I (kt)
\nonumber \\
  & & \times \left[\frac{\xi}{4 \upi t} m_e (t) 
+ \frac{t \xi \rho^{\prime} (t) }{3 \eta }\right]\,\rmd t
\nonumber \\
a_> (r) & = & \frac{a_I (kr)}{k} \int_{r}^{r_{v-}} a_K (kt)
\nonumber \\
  & & \times \left[\frac{\xi}{4 \upi t} m_e (t) 
+ \frac{t \xi \rho^{\prime} (t) }{3 \eta }\right]\,\rmd t\,.
\end{eqnarray}
In the large $z$ limit,
\begin{eqnarray}
a_I (z) & \rightarrow & \frac{\rme^z }{2}
\nonumber \\
a_K (z) & \rightarrow & \rme^{-z} \,,
\end{eqnarray}
so $a_< (r)$ and $a_> (r)$ can be written
\begin{eqnarray}
a_< (r) & = & \int_{r_{\rm min}}^{r} \frac{\rme^{k(t-r)}}{2k}
\left[\frac{\xi}{4 \upi t} m_e (t) 
+ \frac{t \xi \rho^{\prime} (t) }{3 \eta }\right]\,\rmd t
\nonumber \\
a_> (r) & = & \int_{r}^{r_{v-}} \frac{\rme^{k(r-t)}}{2k}
\left[\frac{\xi}{4 \upi t} m_e (t) 
+ \frac{t \xi \rho^{\prime} (t) }{3 \eta }\right]\,\rmd t \,.
\end{eqnarray}
The exponentials are sharply peaked around $t = r$, so we can write
\begin{eqnarray}
a_< (r) & = & \left[\frac{\xi}{4 \upi r} m_e (r) 
+ \frac{r \xi \rho^{\prime} (r) }{3 \eta }\right]
\int_{r_{\rm min}}^{r} \frac{\rme^{k(t-r)}}{2k} \,\rmd t
\nonumber \\
  & = & \frac{1}{2 k^2 } \left[\frac{\xi}{4 \upi r} m_e (r) 
+ \frac{r \xi \rho^{\prime} (r) }{3 \eta }\right] \,.
\end{eqnarray}
In the SS, $m_e (r) = \msun - 4 \upi \rho_b r^3 /3$, and $\rho^{\prime} = 0$,
so
\begin{eqnarray}
a_< (r) & = & \frac{\xi}{8 \upi k^2 r}
\left[ \msun - \frac{4 \upi \rho_b  r^3 }{3} \right] \,.
\end{eqnarray}
$a_> (r)$ will contribute an identical amount, so
\begin{eqnarray}
a (r) & = & \frac{\xi}{4 \upi k^2 r}
\left[ \msun - \frac{4 \upi \rho_b  r^3 }{3} \right]
\nonumber \\
  & = & \frac{\xi \msun}{4 \upi k^2 r}
- \frac{\xi \rho_b r^2 }{3 k^2 } \,,
\end{eqnarray}
in agreement with the PI of (\ref{eq:aPI}). There is no CF in this
approximation, and the Schwarzschild condition is satisfied.

\section{Modelling the density of the Sun}
\label{sec:appa}

We will model the density of the Sun as a Gaussian:
\begin{equation}
\rho(r) = \rho_s \exp \left(-\frac{s_s r^2 }{r_{s}^{2}} \right)\,.
\end{equation}

Requiring that this density decreases by a factor of $10^4 $ as we go from
$r=0$ to $r = r_s$ gives us $s_s = 9.2$.

The enclosed mass out to radius $r$ can be shown to be
\begin{eqnarray}
m_e (r) & = & \frac{4 \upi r_{s}^{3} \rho_s }{s_{s}^{3/2}} E(z)
\hspace{0.25in}{\rm where}
\nonumber \\
z & = & \frac{\sqrt{s_s } r}{r_s} \hspace{0.25in}{\rm and}
\nonumber \\
E(z) & = & -\frac{\sqrt{z}}{2} \rme^{-z}
+ \frac{\sqrt{\upi}}{4} {\rm erf} (z)\,.
\end{eqnarray}

The total mass, $\msun$, is given by
\begin{eqnarray}
\msun & = & \frac{4 \upi r_{s}^{3} \rho_s }{s_{s}^{3/2}}
\int_{0}^{\sqrt{s_{s}}} z^2 \rme^{-z^2} \,\rmd z
\nonumber \\
  & \approx & 0.2 \rho_s r_{s}^{3}\,.
\end{eqnarray}

\section{Estimating the magnitude of
$\bmath{{\mathcal H}_{\rm \MakeLowercase{\displaystyle sun}}}$}
\label{sec:appb}

\begin{eqnarray}
{\mathcal H}_{\rm sun} & = & \frac{\xi}{4 \upi} \int_{r_{\rm min}}^{r_s}
\frac{a_I (kt)}{t} m_e (t)\,\rmd t
\nonumber \\
  & \approx & \frac{\xi k^2}{4 \upi} \int_{r_{\rm min}}^{r_s}
\frac{t}{3} m_e (t)\,\rmd t\,.
\end{eqnarray}
We can let $r_{\rm min} \rightarrow 0$, and take $m_e (r)$ from the previous
section:
\begin{eqnarray}
{\mathcal H}_{\rm sun} & \approx &
\frac{\xi k^2 r_{s}^{3} \rho_{s}}{3 s_{s}^{3/2}}
\int_{0}^{r_s } t E(z) \,\rmd t
\end{eqnarray}
with $z = \sqrt{s_s } t/ r_s $.

We are concerned now only with orders of magnitude. The integral in the
previous equation is ${\rm O} (r_{s}^{2} )$, so
\begin{equation}
{\mathcal H}_{\rm sun} \;\mbox{ is of order }\; \xi \msun k^2 r_{s}^{2} \,.
\end{equation}
The presence of the very small factor $k^2 r_{s}^{2}$ shows that
${\mathcal H}_{\rm sun}$ is negligible in comparison with quantities such as
${\mathcal H}_2 = \xi \msun /(12 \upi)$.

\bsp

\end{document}